# Speckle suppression and companion detection using coherent differential imaging

Michael Bottom,[1] J. Kent Wallace,[2] Randall D. Bartos,[2] J. Chris Shelton,[2] Eugene Serabyn[2]

mbottom@caltech.edu

May 2016

## ABSTRACT

Residual speckles due to aberrations arising from optical errors after the split between the wavefront sensor and the science camera path are the most significant barriers to imaging extrasolar planets. While speckles can be suppressed using the science camera in conjunction with the deformable mirror, this requires knowledge of the phase of the electric field in the focal plane. We describe a method which combines a coronagraph with a simple phase-shifting interferometer to measure and correct speckles in the full focal plane. We demonstrate its initial use on the Stellar Double Coronagraph at the Palomar Observatory. We also describe how the same hardware can be used to distinguish speckles from true companions by measuring the coherence of the optical field in the focal plane. We present results observing the brown dwarf HD 49197b with this technique, demonstrating the ability to detect the presence of a companion even when it is buried in the speckle noise, without the use of any standard "calibration" techniques. We believe this is the first detection of a substellar companion using the coherence properties of light.

*Subject headings:* High contrast imaging, instrumentation: high angular resolution, techniques: interferometric, (stars:) brown dwarfs, Planetary Systems, techniques: image processing

## 1. Introduction

Imaging extrasolar planets is difficult. The typical brightness ratio between planets and stars is a factor of $10^{-4}$ in the best cases of young jovians, and $10^{-10}$ in the case of an Earth twin observed in the visible or near-infrared. Another significant challenge is the small angular separation between the planet and its host star, on the order of 100 milli-arcseconds, only within a factor of a few of the diffraction limit of most space and ground-based telescopes currently in operation.

---

[1]California Institute of Technology, MC 249-17, 1200 E California Blvd, Pasadena, CA 91125, USA

[2]Jet Propulsion Lab, California Institute of Technology, Pasadena, CA 91109, USA

At least three things are required to image extrasolar planets from the ground. First, an adaptive optics system is needed to sharpen the point-spread function and remove substantial blurring of the star and planet light caused by propagation through the turbulent atmosphere of the Earth. Second, a coronagraph is necessary, because even if the optical system is delivering perfect performance, it is necessary to suppress the unavoidable diffraction of the point spread function to allow observations at small angular separations from the star. Finally, some kind of image post-processing is needed to enhance the planet signal-to-noise ratio.

The main purpose of image post-processing is to suppress residual aberrations in the images known as "speckles", which arise from imperfect sensing of optical errors arising in or after the wavefront sensor of the adaptive optics system. These speckles can be tens to thousands of times brighter than any planet, and must be removed in some way. They are by far the most severe limitation for imaging planets, and some major recent breakthroughs in the field have come from new ways of post-processing; LOCI (Lafrenière et al. 2007) and the "KLIP" principal components algorithm (Soummer et al. 2012) being notable examples.

A theoretically superior approach is removing speckles "optically", rather than in post processing, as this reduces photon noise. Speckles are composed of scattered starlight, and thus will interfere with other starlight components. Therefore, for a given speckle, if starlight of the same intensity but opposite phase is directed to the same point in the focal plane, the speckle will vanish. If this is done for every speckle in the focal plane, all speckles will disappear, revealing any underlying companions.

### 1.1. Focal plane wavefront sensing techniques

The first proposed method for focal plane wavefront sensing is called "speckle nulling," described in a pioneering paper by Bordé & Traub (2006). This involves using sinusoidal shapes on the deformable mirror to diffract light to locations of speckles, deriving their phase by changing the phase of the sinusoid on the deformable mirror, then correcting them with the opposite phase. Speckle nulling is a powerful technique as it requires no system model beyond a rough conversion between the sinusoidal amplitude of the deformable mirror to intensity at the focal plane. In stable platforms like controlled high-contrast imaging testbeds, it converges more slowly than other techniques, as the applied correction can amplify the brightness of speckles not being corrected. However, the lack of a model leads to robust performance in non-ideal situations, as was recently demonstrated by Martinache et al. (2014), where substantial contrast improvement was shown on-sky in a low Strehl ratio environment.

A more powerful but model-dependent sensing and correction method is called "electric field conjugation," or EFC (Give'on et al. 2007). The EFC algorithm uses pairs of small actuator displacements to perturb the electric field at the focal plane, in a specified control region. With an accurate model converting deformable mirror actuator displacement to electric field in the focal plane, the phase can be derived and corrected in the control region. The probe shapes are typically quite small in amplitude, as required by the linear formalism, and thus may require somewhat high signal-to-noise in the focal plane to properly measure (Groff et al. 2015). As the governing equations of electric field conjugation are overspecified, typically matrix regularization schemes are used to prevent excessive deformable mirror actuator commands. An extension of electric field conjugation proposed by Pueyo et al. (2011) called "stroke minimization" solves the regularization problem in a logical way, by minimizing the stroke of the deformable mirrors at a given contrast. Electric field conjugation matched with forms of Kalman filtering (Riggs et al. 2016) can provide fast and accurate estimates of both coherent phase errors and incoherent light (for example, from a companion or exozodiacal background). To date, EFC has provided the deepest contrasts measured in controlled environments, at better than $10^9$.

An interesting alternative focal plane estimation method recently proposed by Sauvage et al. (2012), called "coronagraphic focal plane wavefront estimator for exoplanet imaging," or COFFEE, puts known and well-calibrated aberrations on the deformable mirror (typically astigmatism and/or defocus) to introduce diversity at the focal plane, then solves for the phase aberrations in the focal plane using a maximum likelihood estimator. An advantage of COFFEE is that it does not require a full system model, but can use aberrations calibrated by the science camera. Furthermore, rather than using very small probes on the deformable mirror, large aberrations are used, which lead to more dramatic changes to speckle morphology in the focal plane, and hence have higher phase responsivity. On the other hand, unlike electric field conjugation, which would likely work during science observations, it is not clear whether the diversity images, possessing large aberrations, are of much use scientifically. More recent work has decreased the necessary accuracy to which the deformable mirror aberration shapes must be known, increasing performance and making this approach an even more promising method of focal-plane wavefront estimation (Paul et al. 2013). However, to understand the advantages and disadvantages of this method, experimental verification is needed, which has so far been limited.

Speckle nulling, EFC, and COFFEE use the deformable mirror to both measure and correct aberrations. An elegant method proposed by Baudoz et al. (2006) implements a focal plane wavefront sensor optically, using a device called a "self-coherent camera," or SCC. The SCC is implemented by having two small pinholes outside a pupil plane, illuminated by rejected light from a focal-plane coronagraph. The effect of the pinholes in the science camera focal plane is to create fringes over the speckles in the science image, at spatial frequencies finer than the diffraction limit. The fringe offsets and visibilities can be used to measure the phases of the speckles and detect the presence of incoherent light as well. Advantages of this approach are that the measurement is totally static, requires only a single image per iteration, and can be totally integrated into science observations. Disadvantages are that the optical system must be designed around the self coherent camera: the output pupil (and subsequent optics) must be significantly oversized to accommodate the pinholes, and a science camera with high pixel density is needed to sample the fringes.

A more aggressive method of using the coherence of the optical field to discriminate between planets and speckles was described by Guyon (2004), called "synchronous interferometric speckle subtraction," (SISS) where the optical path is separated into two arms, one of which is phase-shifted before they are recombined and reimaged on two separate detectors. The effect of the phase-shifting and recombination is to cause speckles to modulate in time, while the planet signal stays steady. SISS should in principle also be able to measure speckle phases, though this is not discussed in the original paper. Advantages of this technique include the lack of a system model, the high degree of speckle modulation, and especially the ability to work in regimes of quickly changing speckles. Disadvantages are the optomechanical complexity required to implement; perhaps for this reason SISS has never been demonstrated either in a laboratory environment or on-sky.

## 1.2. Coronagraphic phase-shifting interferometry

Here we develop a different approach, first described by Serabyn et al. (2011), which is based on the principles of phase-shifting interferometry (Carré 1966). The method involves using a "reference" beam of light, the core of a broad Airy pattern that covers the entire focal plane, generated by a small mirror at the centre of the output pupil. Pistoning this mirror changes the phase of the reference beam uniformly, as the phase of the electric field at the core of an Airy function is constant up to the first Airy ring. This change in phase changes the intensity of the speckles in the focal plane, due to constructive and destructive interference. It is possible to extract the electric field phase of the underlying speckle field from the measurement of the intensities at the different mirror positions. With this phase knowledge, speckles may be removed by directing light of the same amplitude and opposite phase to the speckle positions using the deformable mirror.

The coronagraphic phase-shifting interferometer has a combination of good features from the approaches discussed above. First, it does not require a good system model, as the reference beam is not created by the deformable mirror but by a small flat mirror. The reference beam is easy to measure experimentally, can be calibrated to good accuracy, does not require oversizing the optical system, and can be commanded to give very large phases, as crosstalk, linearity approximations, and deformable mirror stroke limits are not relevant. Second, the wavefront sensing may be seamlessly integrated into observing, as the intensity modulations of the speckles are not destructive to the image. Thirdly, non-modulation of focal-plane features can be used to discern incoherent light from coherent speckles in a straightforward and model-independent way. Of course, there are disadvantages to this approach. Extra optomechanics are needed, and the reference and speckle fluxes need to be within about two orders of magnitude, requiring some balancing.

This paper will present the theory, and show experimental and on-sky results using this approach. First, we describe describe our implementation of a coronagraphic phase shifting interferometer on the the Stellar Double Coronagraph of Palomar observatory. Second, we present the basic equations governing phase shifting interferometry, which apply exactly in this case. We discuss its implementation as a focal-plane wavefront sensor, including laboratory measurements demonstrating the ability to measure and suppress the electric field in the focal plane. We also present on-sky data of HD 49197, showing how coherence data can be used to detect a faint companion even when it is buried in the speckle aberrations. We conclude by offering an overview of the potential and challenges of this technique, with a view towards future high contrast imaging systems.

## 2. Theory

### 2.1. Phase Shifting Interferometry

Phase-shifting interferometry is a well-known method of extracting spatial information about a wavefront phase by interfering it with a reference wavefront of controlled phase. Following Malacara (2007) we consider an electric field in the focal plane (in this case, the speckle field) described by the equation

$$E_s(x,y) = a_s(x,y)e^{i\phi_s(x,y)} \quad (1)$$

where $a(x,y)$ refers to the corresponding electric field amplitude at $x,y$ (spatial) coordinates in the focal plane and $\phi$ corresponds to the static phase delay of the speckles referenced to some zero point. The subscript $s$ refers to the fact that this is the speckle field. (Quickly oscillating $2\pi ft$ terms in the exponential are omitted for clarity.)

We also consider a "reference" wavefront described by the equation

$$E_r(x,y) = a_r(x,y)e^{i[\phi_r(x,y)-\delta(t)]} \quad (2)$$

The reference wave, in this case, has a static term $\phi_r(x,y)$ and an explicit time dependent phase offset $\delta(t)$ that is controlled externally. For now, we ignore the source of the reference wave, but assume that it is coherent with the speckle field.

The intensity pattern in the focal plane is determined by the squared sum of the electric fields:

$$I(x,y,t) = |E_s(x,y) + E_r(x,y)|^2 \quad (3)$$
$$= a_s(x,y)^2 + a_r(x,y)^2 + 2a_s(x,y)a_r(x,y)\cos[\phi_s(x,y) - \phi_r(x,y) + \delta(t)] \quad (4)$$
$$= I_s + I_r + 2\sqrt{I_s I_r}\cos[\phi_s(x,y) - \phi_r(x,y) + \delta(t)] \quad (5)$$

where the conversion from electric field amplitude to radiant intensity $(a(x,y)^2 = I)$ is used between the second and third line. The terms in the last line of the above equation may be interpreted as a static intensity equal to the individual intensities of the speckle field and reference beam at a given location, plus a modulating term whose intensity depends on the reference beam phase $\delta(t)$. By stepping the reference beam phase $\delta(t)$ in a controlled manner, the intensity of each location $I(x,y,t)$ shows a sinusoidal dependence, with differing amplitudes and phases. The relative phase shift between the different $(x,y)$ locations depends only on $\phi_r$ and $\phi_s$, while the scale factor depends on $\sqrt{(I_r I_s)}$.

One parameter needed to correct the speckle field is the phase term $\phi_s(x,y)$. Consider a flat reference wavefront $\phi_r(x,y) = 0$ (the actual phase will be discussed later in more detail). If the phase of the reference wave is advanced in four quadrature steps of 0, $\lambda/4$, $\lambda/2$, $3\lambda/4 = 0$, $\pi/2$, $\pi$, $3\pi/2$; images $I_1, I_2, I_3, I_4$ may be obtained at each position with the following relations holding:

$$I_1 = I_s + I_r + 2\sqrt{I_s I_r} \cos[\phi_s(x,y)] \tag{6}$$
$$I_2 = I_s + I_r - 2\sqrt{I_s I_r} \sin[\phi_s(x,y)] \tag{7}$$
$$I_3 = I_s + I_r - 2\sqrt{I_s I_r} \cos[\phi_s(x,y)] \tag{8}$$
$$I_4 = I_s + I_r + 2\sqrt{I_s I_r} \sin[\phi_s(x,y)] \tag{9}$$

This is known as the "four-step" algorithm, which yields for the speckle phases:

$$\phi_s(x,y) = \tan^{-1}\left[\frac{I_4 - I_2}{I_1 - I_3}\right] \tag{10}$$

Another quantity of interest is the speckle visibilities $\gamma(x,y)$, equal to the average modulation of the interferogram divided by the mean value

$$\gamma(x,y) = \frac{I_{\max} - I_{\min}}{I_{\max} + I_{\min}} = \frac{2\sqrt{(I_4 - I_2)^2 + (I_1 - I_3)^2}}{I_1 + I_2 + I_3 + I_4} = \frac{2\sqrt{I_s I_r}}{I_s + I_r} \tag{11}$$

The two equations given to extract speckle phases and visibilities are not the only ones possible; it is clear that four steps are actually one more than needed to solve for the three unknowns $I_s, I_r$, and $\phi_s$. On the other hand, numerical instabilities and biases, as well as imperfect measurements of the phase offset $\delta$ can cause three or four steps to give poor estimates of the speckle phases. A discussion of these issues is beyond the scope of this work; see Malacara (2007) for a thorough comparison of different phase-shifting interferometry algorithms.

Phase shifting interferometers can be implemented in various ways, the main differences being in how the reference beam is created and modulated. The simplest is the Twyman-Green interferometer, with the reference beam (for example, a flat mirror) on the moving arm and the test component on the static arm. Another implementation is the Mach-Zehnder configuration, with one of the 45 degree mirrors actuated. Usually, the source used to illuminate the optics is a bright, coherent laser, which is quite different from the situation where the star itself provides the flux.

### 2.2. Coronagraphic phase shifting interferometer design

Phase shifting interferometry requires a coherent reference wave to interfere with the light distribution of interest, and this reference wave needs to be bright enough so that the fringe visibilities are measurable. When used with a coronagraph, this reference beam must consist of starlight, and its intensity, size and shape in the focal plane, and scattered light must be controlled.

In this case, the reference wave was generated from residual starlight from the coronagraph optics which is usually blocked. Coronagraphs are good at redistributing light around subsequent image and pupil planes in non-intuitive ways, and the implementation of a coronagraphic phase-shifting system thus requires design around a particular architecture. The particular coronagraph design we use is called a "dual-vortex" coronagraph, which has two internal focal planes and two internal pupil planes.

We first describe the operation of the dual-vortex coronagraph design in slightly more detail, with the next section covering the optomechanical setup. The top row of Figure 1 shows the light distribution of different pupil and focal planes through the instrument, and the bottom row shows the optical elements generating these light distributions, with the optical vortices represented by slotted circles. For an obscured input pupil (top row, leftmost panel), the effect of a single vortex and a Lyot stop leaves a large amount of residual diffraction in the following focal plane (top row, 3rd panel from right). The action of the second vortex is to take the residual starlight and concentrate much of it inside the geometric shadow of the secondary (2nd panel from right).

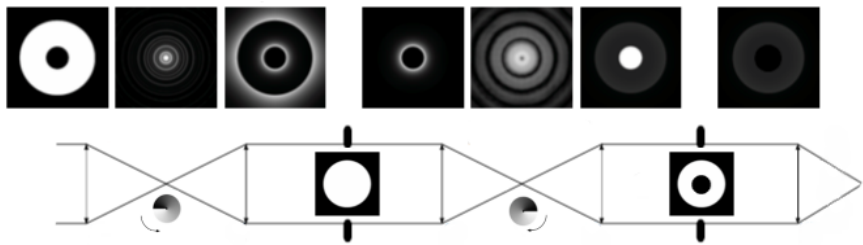

Fig. 1.—: Schematic of the dual-vortex coronagraph. The first vortex leaves a residual halo of light (4th panel from left) which is moved behind the pupil by the second vortex (2nd panel from right). In the second pupil plane, this light can be blocked, creating an effective conventional coronagraph. Alternatively, it can be picked off and used as a "reference" beam, as we do here with a phase-shifting mirror, described below. This figure originally appeared in Bottom et al. (2016).

Ignoring phase terms, if the uniformly illuminated input pupil (the leftmost panel) has an electric field distribution of

$$E(r) = \begin{cases} 0 & r < r_0 \\ 1 & r_0 < r < R \\ 0 & r > R \end{cases}$$

then the electric field distribution just before the second Lyot stop (penultimate rightmost panel)

will be

$$E(r) = \begin{cases} (r_0/R)^2 - 1 & r < r_0 \\ (r_0/R)^2 & r_0 < r < R \\ 0 & r > R \end{cases}$$

where $r_0, R$ are the radius of the secondary and primary mirror, respectively (Mawet et al. 2011). The typical optic at the second pupil would be a Lyot stop blocking the bright core of light. Instead, we replace this optic with a phase-shifting mirror to use a portion of this bright core of light as the reference beam.

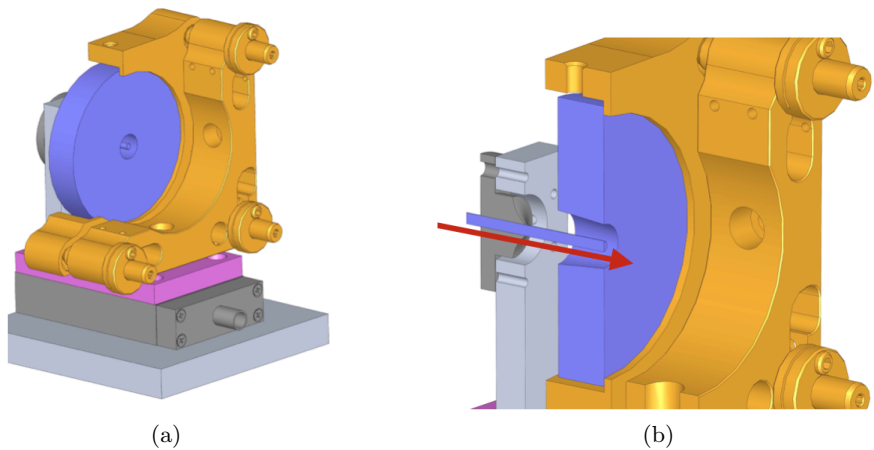

(a)          (b)

Fig. 2.—: (a) Mechanical drawing of the phase shifter mount. The piezoelectric flexure stage is shown in dark grey, the interface to the annular mirror mount is in pink, and the annular mirror mount is in yellow. The annular mirror mount is actuated on the three axes, but the actuators are omitted for clarity. The mirrors are shown in blue. (b) Cutaway of the phase-shifter assembly, showing the hardware to align the rod to the mirror and the direction of travel of the rod. The mechanism is installed at the "Lyot Stop 2" location in Figure 4.

## 3. Optomechanical implementation

### 3.1. Stellar Double Coronagraph

The Stellar Double Coronagraph (SDC) is a JPL-developed instrument designed for high-contrast imaging of close-in companions to stars. It supports multiple observing modes, including single and dual vortex coronagraphs (Mawet et al. 2011), ring-apodized coronagraphs (Mawet et al. 2013), and a self-coherent camera (Baudoz et al. 2006) . In typical observing situations, it has an inner working angle of approximately $1\lambda/D$, or 90 mas in K-band (2.2 $\mu$m) behind the 5 m Hale telescope. It is installed between the P3K adaptive optics system (Dekany et al. 2013) and the near-IR imager PHARO (Hayward et al. 2001). Figure 4 shows a layout of the instrument. A thorough description of the instrument and its observing modes may be found in Bottom et al. (2016).

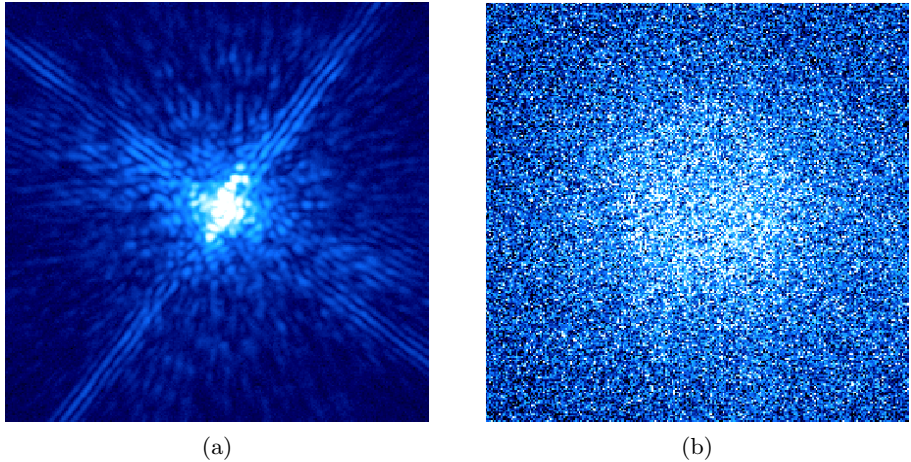

(a)            (b)

Fig. 3.—: (a) The point spread function of the combined outer mirror and phaseshifting rod. (b) The point-spread function of the inner phase-shifting rod only. The images are the same size, with each edge spanning 6.25 arc-seconds (250 pixels). The images are not at the same scale or exposure time, but have been individually stretched to bring out the speckle field and reference wave extent.

The primary observing mode of the SDC uses two optical vortices in series. This has the effect of simultaneously diffracting starlight out of the pupil of the instrument and partially correcting for the secondary obscuration of the telescope. (Secondary obscurations and other pupil artifacts can cause serious contrast and inner working angle degradations, compromising performance on most coronagraphic designs, and must be accommodated in some way.) Figure 1 shows the focal and pupil planes of the instrument, demonstrating the effects of the optical vortices on the input starlight.

We implemented a phase-shifting interferometer by replacing the conventional flat mirror in the second Lyot plane (labelled "Lyot stop 2" in Figure 4) with an annular mirror. A piezoelectric flexure stage (Physik Instrumente P-752) underneath the annular mirror mount drives the small inner mirror, which is centred in the annular mirror. The hole in the mirror is beveled and the collar holding the inner mirror is canted to minimize scattered light from the remaining bright core of light at the centre of the pupil. The piezoelectric stage and controller have a resolution of a few nanometers and a range of 30 $\mu$m, corresponding to a precision of a few thousandths of a wave and a dynamic range of nearly 30 waves at the centre of $K_s$ band, 2.15$\mu$m. A mechanical drawing of the apparatus is shown in Figure 2. Images of the point-spread function of the outer annular mirror and inner mirror (the reference beam) are shown in Figure 3.

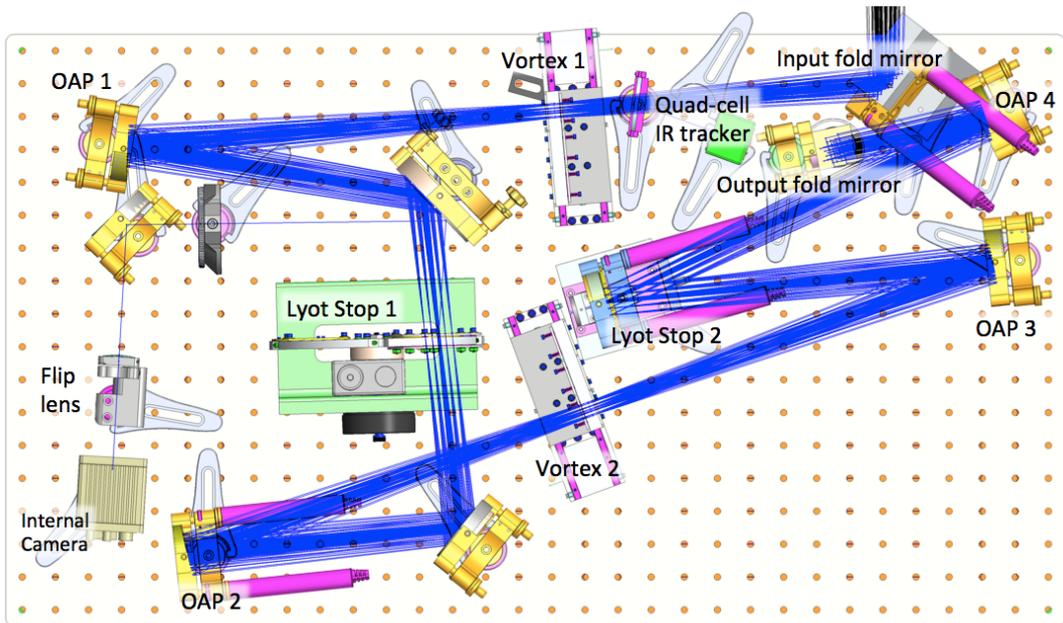

Fig. 4.—: The optomechanical layout of the SDC. Following the input beam from the top right of the figure: first fold mirror, dichroic beam splitter, linear coronagraphic slide, off-axis paraboloid, fold mirror, Lyot plane, fold mirror, off-axis paraboloid, linear coronagraphic slide, off-axis paraboloid, second Lyot plane, off-axis paraboloid, fold mirror. The infrared tracker is the green square. The image and pupil viewing camera and lenses are shown on the left, directly below the first off-axis paraboloid. In this orientation, the output beam to the infrared imager PHARO exits downward into the page. The phase-shifter assembly is at the second Lyot plane, labelled "Lyot Stop 2" in the image. This figure originally appeared in Bottom et al. (2016).

### 3.2. Initial setup and alignment

We first performed a number of tests of coronagraphic phase shifting interferometry in a laboratory setting, using the P3K adaptive optics system, the SDC, and the near-infrared camera PHARO. All tests were performed in the infrared $K_s$ band, with a central wavelength of $\lambda = 2.145$ $\mu$m and a bandwidth of 1.99 - 2.3 $\mu$m ($\sim$15%), identical to our on-sky observing configuration. The tests were performed using the internal white light fiber of the adaptive optics system as an input source, and a temporary mask on the deformable mirror to simulate the pupil of the Hale telescope while in the lab.

The initial setup consisted of separately imaging the point-spread function of the annular mirror and phase-shifting mirror on the detector, then coaligning their centres by tilting the actuated annular mirror. Images of the two point-spread functions may be seen in Figure 3. The coherence length of this interferometer can be approximated by $\lambda \cdot (\lambda/\Delta\lambda) = 4.7$ $\mu$m, which is easily within the range and precision of the actuators on the optics. Co-phasing the mirrors was accomplished by pistoning the annular mirror until the white-light fringe (the point of highest fringe contrast, corresponding to zero path length difference) was acquired. While acquiring the white light fringe is not a necessary condition for phase measurements, it is helpful for obtaining the greatest fringe contrast, so a single long scan of the phase-shifting mirror was used to determine the white-light fringe position. An example of such a scan is shown in Figure 5.

Finally, the two optical vortices were coaligned with the input AO source, creating the correct output pupil with the bright core at the position of the phase-shifting mirror. A quad-cell tracker internal to the SDC measured slow drifts off the first vortex, sending corrections to the adaptive optics system in a separate loop operating outside the normal 2kHz closed-loop operation. This was important because coronagraph/PSF misalignments can amplify the residual diffraction in the image plane, increasing the speckle brightnesses and creating measurement error.

## 4. Focal plane wavefront sensing via phase shifting interferometry

### 4.1. Phase accuracy from random errors

It is instructive to consider the accuracy to which phase may be measured in the interferometric setup. Using the four-step algorithm described earlier, it is relatively straightforward to propagate the uncertainties in the phase:

$$\phi_s(x, y) = \tan^{-1}\left[\frac{I_4 - I_2}{I_1 - I_3}\right] \tag{12}$$

$$\Rightarrow \sigma_\phi^2 = \sum_{i=1}^{4} \sigma_{I_i}^2 \left(\frac{\partial \phi}{\partial I_i}\right)^2 = \frac{(I_s + I_r)}{4 I_s I_r} \tag{13}$$

$$\Rightarrow \sigma_\phi = \frac{\sqrt{I_s + I_r}}{2\sqrt{I_s I_r}} \tag{14}$$

which may be rewritten in terms of the mean image and visibility as

$$\sigma_\phi = \frac{1}{\gamma\sqrt{2\langle I \rangle}} \tag{15}$$

The derivation follows after some algebra and assuming Poissonian statistics, i.e. $\sigma_I^2 = I$ (see the Appendix for details). For a fixed flux $F = I_s + I_r$, the minimal value of $\sigma_\phi$ occurs when $I_s = I_r = F/2$, as expected. Note that the equality of speckle and reference beam intensity is exactly the case in speckle nulling, by construction, where the "probe" speckles are matched to the intensity of the offending speckles. We can also see the penalty we pay by using a non-matched reference beam. For example, if both the reference and speckle are 1000 counts, the uncertainty is only about 3 degrees, whereas if the reference beam is 100 times fainter than the speckle (ie, 10 counts), the expected uncertainty in phase is on the order of 20 degrees. Of course, if the errors are random, the algorithm should still converge, albeit perhaps more slowly than desired.

### 4.2. Phase accuracy from systematic errors

Unfortunately, we were not limited solely by random error, as we discovered one additional complication in accurate phase retrieval. In the ideal case, the phase of the reference beam should be flat over the entire focal plane, and at a constant phase if the mirror is centred perfectly. A decentration of the pistoning mirror in the pupil plane leads to a phase tilt of the reference beam Airy pattern across the focal plane, which may be removed during data processing. However, it became apparent there was a static (up to a constant offset) phase gradient over the image. To determine the precise shape of the phase gradient, we generated spots of zero phase on the deformable mirror, i.e. cosine waves with a peak centred on the mirror. (The choice of zero DM phase for the reference spots simplifies the determination of the phase gradient, as the electric field phase will be the same everywhere, unlike a choice of 45 degrees, where half the focal plane would have electric field phase of +45 and half would have spot phases of -45.) We then performed a measurement of the phase using the same four-step algorithm. We extracted the phase measurements from each calibration spot and generated a surface plot of the phase gradient (see Figure 7). An example of measurements of phase in the focal plane, and their interpretation, is presented in Figure 6.

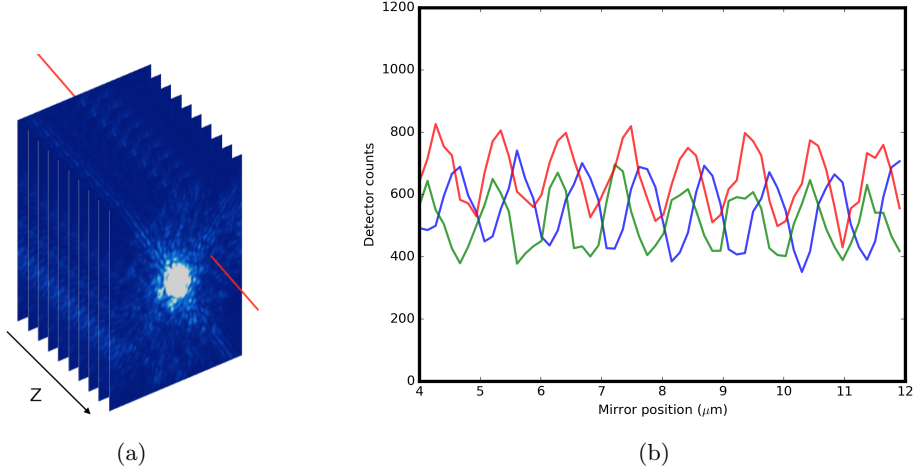

(a)             (b)

Fig. 5.—: (a) An image is taken at each pistoning mirror position (z). The red line shows a cut through a particular pixel, at different mirror positions. (b) Raw data of intensity (in counts) vs mirror position for three different pixels of similar intensity in the image plane. The relative phases between the three different waves correspond to differences in electric field phase.

The phase gradient surface was very well described by the sum of a plane and a paraboloid $(Ax^2 + By^2 + Cxy + Dx + Ey + F)$, suggesting that it was caused by a decentration of the phase-shifting mirror and a focus term. The focus term could be due to some optical power on the pistoning mirror. The residual phase error from the fit was typically 5-10 degrees, which sets the maximum phase accuracy attainable here. The phase gradient was determined to be quite stable, with no significant changes observed over the course of the experiment.

However, it was found that the zeropoint (that is, the offset $F$ in the paraboloid equation) of the phase was not stable to this accuracy level over periods longer than 1-2 hours, which we attributed to thermal and mechanical drifts between the rod and annular mirror beams. The zeropoint of the phase is critical, of course, for having a consistent correction applied by the deformable mirror. A single culprit for this drift was not found, but it is likely that the mechanical coalignment stability was at fault.[1] The solution to this was to simply rerun the calibration of the phase gradient every 1-2 hours or so.

### 4.3. Chromatic effects

Chromatic effects will of course appear when carrying out phase measurements with broadband light. We first note that a single sinusoidal ripple on the deformable mirror (DM) will produce a pair of focal plane speckles located on opposite sides of the center, at a radius of $k\lambda/D$, where $k$ is the DM ripples spatial frequency. Because the speckle position is proportional to wavelength, speckles far from the central position will tend to elongate in broadband light. Moreover, the phase of the focal plane speckles is given directly by the phase of the DM ripple, i.e., by the phase shift, $\phi$, of the surface cosine wave with respect to the center of the mirror (with the pair of focal plane speckles having opposite phases of $\pm\phi$). While the speckle position is thus wavelength dependent, *the phase is not*, as there is only a single ripple on the DM in this example, with a single, unique phase, that applies to all wavelengths.

---

[1] we calculated that 30 nm of path error corresponds to 10 degrees of phase accuracy

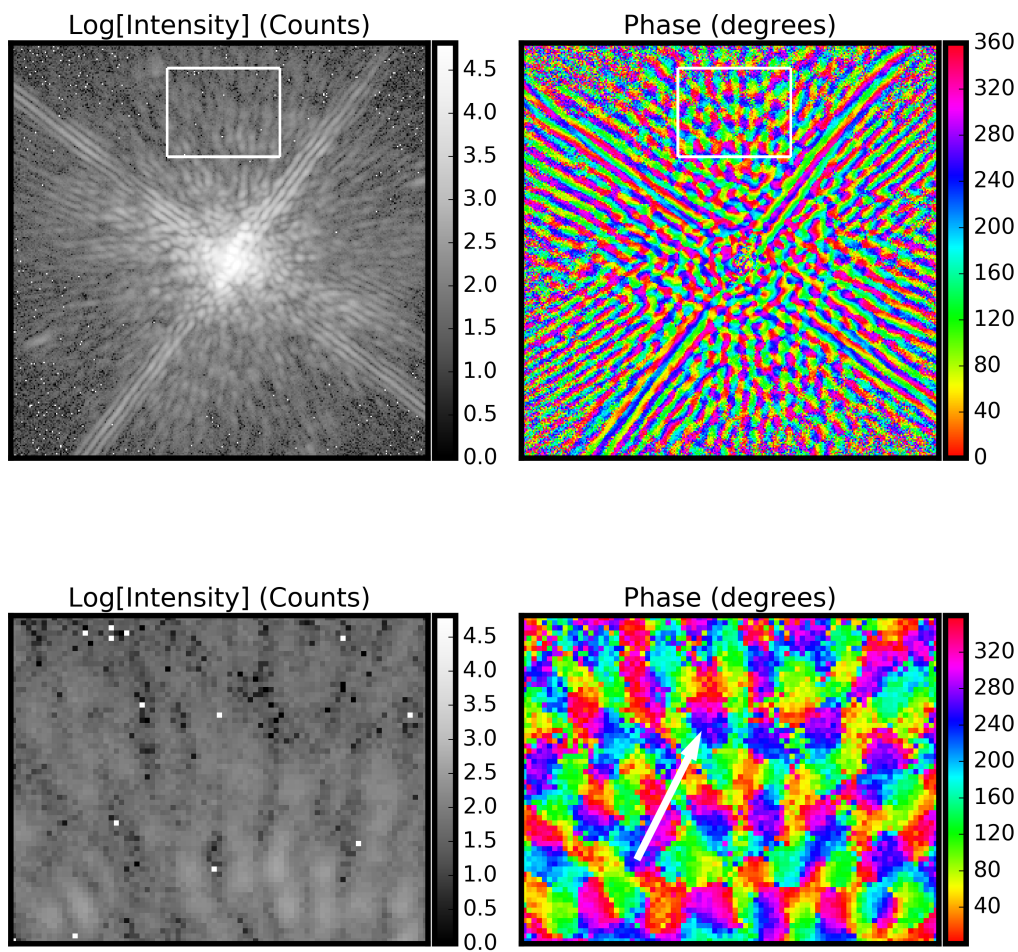

Fig. 6.—: Phase sensing in the focal plane. (Top row) The left-hand plot shows the intensity measurements in the focal plane, with diffraction spikes clearly visible. This measure of intensity is a typical image shown on an camera. On the right, the corresponding phase map is shown. (Bottom row) The outlined region in the top plot, magnified to show detail. Speckles and their Airy rings are more easily identified in the phase plot, with a 180 degree phase shift between the cores and first Airy rings, as indicated by the arrow.

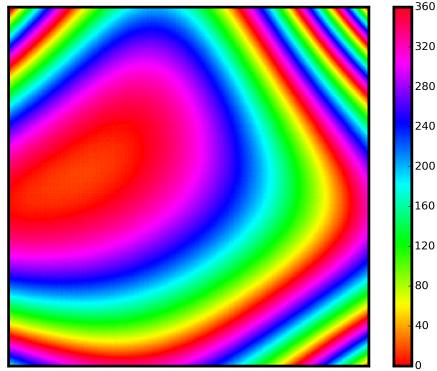

Fig. 7.—: The derived map of the phase gradient over the same area as in Figure 6.

However, in our measurement scheme, the optical effect of the phase-shifting mirror is different from that of the DM. In particular, moving the phase-shifting rod yields a wavelength-dependent phase shift, as the motion is along the direction of propagation. When using broadband light, the $\delta(t)$ term in Equation is thus wavelength dependent, as $\delta(t) = 2z \cdot 2\pi/\lambda$, where $z$ is the mirror position. Equation 10, as given, did not include this effect, and so a chromatic error will result from the application of it to broadband speckle phases. Considering an elongated phase speckle generated from light spanning the wavelengths $\lambda_{\text{lo}}$ to $\lambda_{\text{hi}}$, the phase measurement will only be accurate at the wavelength for which the step size is indeed a quarter of a wavelength, i.e., near the center of the speckle. The net effect will thus be that the measured phase will shift regularly across the speckle.

As the generalization of 4.3 to include this effect is non-linear, we chose to simulate this chromatic effect numerically, by computing equations 4.3 and 10 over a range of bandwidths and phases. The net phase shift across a speckle was found to scale roughly linearly with the bandwidth for small bandwidths; for example, 15% bandwidth light leads to a mean phase shift of about 19 degrees across a speckle of random phase, ranging from 13 to 25 degrees, depending on the speckle phase. For light of only 3.5% bandwidth, the mean phase shift was 4.7 degrees, with a minimum of 3.15 and a maximum of 6.28. The calculated phase shift was found to oscillate sinusoidally as a function of speckle phase, with an amplitude of about 3 degrees/cycle in the latter case.

To test the validity of the calculations, we generated speckles of 0 degrees of phase in 15% broadband light ($K_s$) using the DM, and then measured their phases with our phase-shifting interferometer. Our calculations gave an expected phase shift of 25 degrees across the speckles for this case (decreasing outward). To compare to this, we extracted the angular average of the measured phase vs the radial position for the inner 16 speckles. A clear gradient in phase is visible (Figure 8), of the correct sign, and with a net phase shift across the speckle of about 15 degrees. This is less than predicted, likely due to under-sampling by the camera—with only a few pixels across the core of the PSF, the measured phase would tend to get smoothed out, an effect which can also be modelled. The small chromatic phase shift that is observed can thus be modelled and removed when needed, leading to improved accuracy. For more modest bandwidths, the small phase shifts can be ignored.

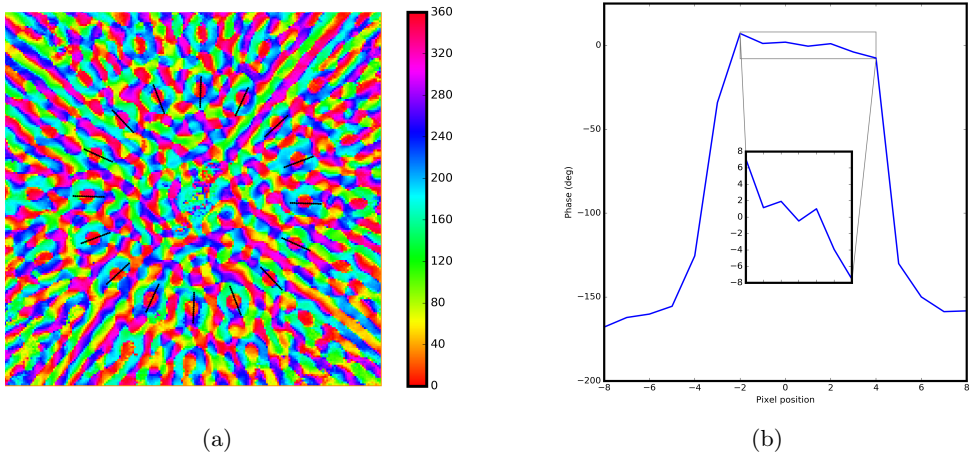

Fig. 8.—: (a) The measured phase of the speckle field in the focal plane. 32 speckles of zero phase generated by the deformable mirror are visible. The lines over which the phases of the 16 inner speckles are measured in the right plot are indicated in black. (b) The angularly averaged phase over the 16 speckles, shown as a function of radial pixel position centered on the speckle. The plot extends from either side of the Airy rings, which are ∼180 degrees out of phase from the cores, as expected. In the cores of the speckles, the predicted decreasing phase gradient is visible, shown in the inset.

### 4.4. Dark hole generation

An example of a dark hole generated using the phase-shifting approach is shown in Figure 9, improving contrast by a factor of 3-4 over the control region. While the dark hole is evident, there were shortcomings in this implementation which prevented this technique from working to its full potential. First, despite the phases and amplitudes of each pixel being measured over the full focal plane, with limited time available, we were not able to implement a full-field solution on the deformable mirror–we just proceeded in the same manner as speckle nulling, picking out bright points and using the average measured phases to generate sinusoids on the deformable mirror to cancel them. This unfortunately handicaps one of the main strengths of this technique, which is the full-field measurement and broadband chromatic correction. Second, the phase gradient was measured using only 32 calibration spots over the field, and in retrospect, at somewhat worse signal-to-noise as we would have liked. This left systematic errors at the 10-15 degree level, which directly propagate into the measured phases, driving the convergence to a slightly wrong setpoint. As a result, the performance of the phase shifting approach in this initial trial was not as good as the performance produced on the same speckle field with speckle nulling, with the dark hole contrast being worse by a factor of two to four, and the convergence being somewhat slower.

## 5. Coherence-based companion detection

### 5.1. Principles

Similar to other high-contrast imaging instruments, the typical observing strategy used with SDC is driven by the need to remove speckle noise in the final, post-processed image. The Hale telescope has an equatorial mount, so the sky plane is fixed with respect to the detector plane. This does not allow for techniques like angular differential imaging (Marois et al. 2006), where sky rotation can be used to separate any (rotating) companions from the static speckles. The strategy we normally use during observations, called "reference differential imaging" quickly alternates between the target of interest and a nearby star of similar $V$ and $K$ apparent magnitudes. The proximity of the stars guarantees that flexure in the optical path stays relatively constant between target and reference, leading to very similar speckle patterns; the same $V$ and $K$ ensure the same adaptive optics performance and flux on the detector plane, respectively. During data reduction, it is possible to then remove the speckles surrounding the target star using the reference star PSF, using either classic point-spread function subtraction or more involved methods like LOCI (Lafrenière et al. 2007) and KLIP (Soummer et al. 2012).

During the observing sequence, we repeated the standard procedure but added one modification, where we synchronized the exposures to the pistoning mirror position, which advanced in steps of $\lambda/8$. Each image then had a particular mirror phase associated with it, which could be used in post-processing to generate "coherence maps" over sequences of images using the visibility formula (Equation 11).

Because the visibility depends on the geometric mean of the speckle and reference wave, in the case of a speckle with the same brightness as the reference wave ($I_s = I_r$), the visibility will be 100%, in the case of a speckle one thousand times brighter ($I_s \gg I_r$), the speckle will modulate at about 6%, and for one ten thousand times brighter the modulation will be about 2%. Ignoring speckles, what about a companion in the image? Since the light from the star is incoherent with the light from the companion, there will be no interference, so $I_1 = I_2 = I_3 = I_4$; $\gamma = 0$. Of course, companion and speckle light will overlap, so what will actually be detected is a region of lower local visibility in the image at the location of the companion. Thus it is possible in *principle* to detect a companion by looking for local minima in the visibility. However, because the visibility depends on the relative brightness of the speckle and reference beam, while the reference beam is locally uniform in intensity, speckles will vary in different parts of the image. For example, speckles tend to be brighter nearer to the star (due to more errors in optical figures at low spatial frequencies), so visibility is expected to be lower there. Therefore, care must be exercised, as a brighter-than-average speckle could also be responsible.

There are a few ways to resolve this brighter-than-average-speckle/companion ambiguity, which require some additional assumptions or measurements. We note that without any other information, it is impossible to resolve this, as the equations in Section 2.1 are unchanged upon exchange of $I_s$ and $I_r$.

In the discussions that follow, we will refer to two derived observables which we repeat here, for clarity:

$$\gamma = \frac{2\sqrt{(I_4 - I_2)^2 + (I_3 - I_1)^2}}{I_1 + I_2 + I_3 + I_4} \tag{16}$$

$$\langle I \rangle = \frac{I_1 + I_2 + I_3 + I_4}{4} \tag{17}$$

where the $I_j$'s are the individual images taken at each phase step.

One way to proceed is when the shape of the reference wave is known up to a multiplicative constant in intensity, so that $I_r[x,y] = cI_{rn}[x,y]$, where $I_{rn}$ is some normalized version of the reference wave. This reference wave shape may be computed from physical optics simulations, or measured, see for example Figure 3. In that case, the following holds:

$$\langle I \rangle = I_s + cI_{rn} + I_p \tag{18}$$

$$\gamma = \frac{2\sqrt{cI_{rn}I_s}}{cI_{rn} + I_s + I_p} \tag{19}$$

$$\Rightarrow I_p = \langle I \rangle - cI_{rn} - \frac{\gamma^2 \langle I \rangle^2}{4cI_{rn}} \tag{20}$$

If it is not possible to measure the shape of the reference wave, it can be approximated as being a multiple of the speckle field, so, $I_r[x,y] = qI_s[x,y]$, where $q$ is a *scalar*. This approximation is somewhat valid, because the speckle intensities are amplified by underlying diffraction, and the reference wave is shaped approximately like this diffraction pattern, with the intensity falling off at a similar rate as to the speckles. (This situation would apply exactly in the case of a two-beam interferometer, for example, where the images are recombined after a static phase shift is applied to one of them, as first described by Guyon (2004)). Therefore,

$$\langle I \rangle = (1+q)I_s + I_p \tag{21}$$

$$\gamma = \frac{2\sqrt{q}I_s}{(I+q)I_s + I_p} \tag{22}$$

$$\Rightarrow I_p = \langle I \rangle \left(1 - \frac{1+q}{2\sqrt{q}}\gamma\right) \tag{23}$$

In both of the previous two cases, the constants $c$ and $q$ must be determined. In principle, they may be found by using a priori knowledge of the optical system–for example, in the case of the SDC, the intensity of the electric field at the centre of the second pupil (in Figure 1) combined with the width of the annular mirror predicts a particular intensity in the focal plane. However, the predictive approach is dangerous for on-sky observing, because the illumination of the optics may change somewhat during the course of an exposure sequence, especially due to tip/tilt error. It is safer to use the optical setup to predict a reasonable starting point for these parameters and experiment with the precise values in post-processing. Regardless, these two methods give the planet light at the cost of one further parameter.

A way to proceed that does not rely on "free" parameters is to observe a nearby reference star, as usual. Apart from the pixels corresponding to the planet position, the reference star will have the same visibility map as the target star, regardless of their relative brightness. This is because the value of $\gamma$ depends on on the relative values of $I_s$ and $I_r$, and not on their absolutes—scaling each up by the same factor does not affect the coherence, and the relative brightness of the reference beam $I_r$ is set by the optics. Letting the visibility map of the target and reference star be $\gamma_T$ and $\gamma_R$, respectively, we find

$$\gamma_T = \frac{2\sqrt{I_{s,T}I_{r,T}}}{I_{s,T} + I_{r,T} + I_p} \tag{24}$$

$$\gamma_R = \frac{2\sqrt{I_{s,R}I_{r,R}}}{I_{s,R} + I_{r,R}} \tag{25}$$

$$\tag{26}$$

where we have explicitly included the light from a planet or other companion ($I_p$) in the target image. It then follows that

$$\frac{1}{\gamma_T} - \frac{1}{\gamma_R} = \frac{I_p + I_{r,T} + I_{s,T}}{2\sqrt{I_{r,T}I_{s,T}}} - \frac{I_{r,R} + I_{s,R}}{2\sqrt{I_{r,R}I_{s,R}}} \tag{27}$$

$$= \frac{I_p}{2\sqrt{I_{r,T}I_{s,T}}} = \frac{I_p}{\gamma_T(I_{r,T} + I_{s,T} + I_p)} = \frac{I_p}{\gamma_T \langle I \rangle} \tag{28}$$

$$I_p = \left(1 - \frac{\gamma_T}{\gamma_R}\right) \langle I \rangle \tag{29}$$

where the term $\langle I \rangle$ refers to the mean value of each pixel over one phase cycle; that is, $(I_1 + I_2 + I_3 + I_4)/4$. While the use of a reference star would seem to partially negate the advantages of using coherence, as such a star is not usually needed, it is still possible to get extra information or amplify the incoherent signal in this way.

In summary, coherence data may be used to discover the presence of companions in a relatively straightforward manner. Without any assumptions, a companion may be inferred as a region of locally reduced visibility, but this is insecure, as unusually bright speckles can also cause locally reduced visibility. To break this uncertainty, either knowledge of the reference beam shape or assumptions of the relative falloff may be used to solve for the incoherent component of the image, at the cost of an additional fitted parameter. Finally, observing a reference star can break this degeneracy and solve for the incoherent light without any further assumptions or parameters.

### 5.2. Combination with image post-processing

Image processing is an important aspect of high-contrast imaging, and one must always consider how different observing strategies interplay with post-processing to affect signal-to-noise of any detected companion in the final data product. We briefly provide an overview of the popular KLIP algorithm to motivate the discussion, while noting that most of what follows is also applicable to LOCI. The KLIP principle components algorithm requires a "reference" set of images containing "identical" speckles, but not the companion, from which it generates a low-dimensional subspace. The "target" images are then projected onto this low dimensional subspace, and the projected images are subtracted from the original target frames. The companion light will then be left over in the final images. Usually, the reference images are taken from nearby stars, which causes the speckle pattern to shift somewhat, as the gravity vector in the optical system changes. They may be also taken from the same star if the telescope permits sky rotation with respect to the optical system, as in the case of alt-az designs; this is the principle behind "angular differential imaging," where the companion rotates with respect to the speckle field. The angular differential imaging approach has the advantage that the speckles barely change, but requires large amounts of observing time for substantial field rotation, and is much less effective at close inner working angles where companions typically like to exist.

In our case, we can combine the coherence modulation with KLIP seamlessly, simply by com-puting the principle components using the coherent light in the image. For example, consider Equation 20, which gives the planet light as a function of the mean intensity, visibility, reference beam shape, and a scaling parameter. As written, it can be interpreted as "PSF subtraction," where the PSF is given by the coherent light in the image. Instead of subtracting, we can instead use the coherent part of the image–that is, $I_s = \gamma^2 \langle I \rangle^2 / 4cI_{rn}$–to compute the principle components. In practice, this involves calculating $I_s$ for each set of four images, then using the set of all the $I_s$'s to generate the principle components.

This is quite useful for two reasons. First, in the absence of noise, the coherent parts of the image will match the speckles very well, as no changes in the optical system will have occurred, unlike when using a reference star. Said another way, the principle components are calculated from the target frames themselves, just with the companion light missing, since it's incoherent. Second, the "scaling" parameter $c$ essentially drops out of the calculation, since principle components analysis will return the same low-dimensional subset regardless of whether the data is scaled by a constant factor. There is one slight caveat to this, which is that $c$ actually appears twice in the $\gamma$ term, in the numerator as a constant multiplicative factor, and in the denominator as part of a quantity $I_s + cI_{rn} + I_p$. The denominator term is dominated by $I_s$, so the contribution of the $c$ can be safely ignored.

Nothing is free, of course, and the price to be paid for using the coherence data to generate the principle components is that the calculated components are noisier than those that could be generated from an appropriately chosen reference star, at least in this particular optical setup. However, the ability to generate companion-free components without using a reference star allows much more observing time to be spent on the target star, which is beneficial for observing efficiency.

### 5.3. Demonstration: HD 49197b

To demonstrate the feasibility of the techniques described above, we observed the star HD 49197 on 22 November 2015 using the phase-shifting interferometer mode on SDC. The star has a known brown-dwarf companion separated by 0".95 with a brightness difference of $\Delta K_s = 8.22 \pm 0.11$, approximately 2000 times fainter than the primary (Metchev & Hillenbrand 2004). The star was chosen because the companion was at approximately the same brightness as the surrounding speckle field, maximizing the expected coherence differences, and also because the companion is bright enough to be detectable in a reasonable amount of observing time. Measurements of the companion separation, position angle, contrast, and a detection of orbital motion are presented in the Appendix.

Observing conditions were typical, with the adaptive optics system delivering a Strehl ratio of about 70-75%. We synchronized the exposures of the infrared camera PHARO with the phase shifting mirror, so that the phase of the reference beam was recorded with each image taken. We took 180 images of the target frames, and 180 images of the reference star, with the mirror modulating in steps of $\lambda/8$ from 0 to $1.25\lambda$. This allows for two sets of independent visibility measurements per mirror phase cycle, that is, interleaved steps of $\lambda/4$.

The data analysis steps generally proceeded as described in the previous section, with a few modifications that we list here. First, the individual frames were preprocessed with bad pixel interpolation and flat field correction. Four frames out of the 360 suffered from a known error in the PHARO camera where all the pixels in the image are read out at very large negative values; this required us to discard those four interferometer sequences (by sequence, we mean each four-step measurement where the interferometer moves by $\lambda/4$ per step). As visibility measurements are very sensitive to changes in total brightness from image to image, we selected the most useful data by calculating the variance of the total flux in each image sequence, and discarding the half with the highest variance (the discarded data is perfectly fine for normal image reduction, but not for extracting coherence information). At the end of this preconditioning, we were left with nineteen sequences of four images each, for a total of 76 images.

For each sequence, we calculated the mean image and visibilities, and use these to create the incoherent intensity map according to Equation 29. (The shape of the reference wave had already been measured in the lab, and is shown in Figure 3). The 19 incoherent intensity maps are then median combined for a final incoherent intensity image. We also present an image of the incoherent signal-to-noise ratio per pixel, which is divided by the shot noise contribution per pixel, estimated as the square root of the mean intensity. We "regularized" two regions of the image which had unphysical values; the outer edge where the reference beam did not reach, and the vertical diffraction spike. These areas had very large negative incoherent intensity fluxes, as the visibility noise was too high. These regions are fairly easy to identify in the incoherent signal-to-noise ratio image, Figure 10, at the outer edges of the image and at position angles of 0 and 180.

Analysis of the coherence modulation data clearly revealed the brown dwarf, despite the imperfect observing conditions (see Figure 10 b, c). Perhaps the most interesting result was that in the averaged target frames (Figure 10 a), the companion was not readily detectable above the speckle noise floor, while using the *same* image data and visibility information, the brown dwarf appears as a significant local maximum, without using any reference star frames at all. This essentially amounts to "orthogonal" evidence of the presence of a companion, pointing to a potentially useful line of attack against speckle noise limits in companion detection.

We also performed the reduction described in Section 5.2, where we calculated the principle components using light from the coherent part of the image, then use those to optimally subtract the point-spread function of the star. The KLIP approach removed starlight efficiently and clearly revealed the companion, but created a great deal of high-frequency noise due to the imperfect extraction of the components, which we elected not to remove. To perform the PCA reduction, we used an annular-zone algorithm that had two components per 100 pixels. This image is presented in Figure 10 c.

It is interesting to note that using coherence seems to be an efficient method of removing starlight without using a reference star. The mean image uses 178 frames and achieves a SNR of about 5. The incoherent intensity image, using the phasing information to extract the planet signal, uses 76 frames and achieves a SNR of about 7.5. Using a reference star with classic (annular) PSF subtraction[2], and matching the same number of total images, a SNR of 9.3 is achieved. This is only about 25% better than not using a reference star but using the coherence information, and the same SNR as using principle components analysis with a reference star and equal number of frames (38 target and 38 reference). Using the "coherent" principal components approach without a reference star, where the components are calculated from the coherent intensity only, the SNR is somewhat surprisingly only 5.35 , despite the companion being cleanly separated (Figure 10 c). This appears to be due to the high-frequency noise introduced in the principal components, which is evident in the image, and we elected not to remove. In particular, large negative values near the core of the PSF of the companion dominate the noise term. When calculating PCA conventionally using all the frames from the reference star to build the components (image not show), and all the frames from the target star, we managed a SNR of about 20. A summary of these results is presented in Table 1. While some caution is urged, as the effects on SNR from data reduction algorithms and physically distinct observing styles are somewhat difficult to disentangle, the good performance of coherence modulation on this first test is encouraging.

| Method | Target images | Reference images | SNR | Note |
|---|---|---|---|---|
| Mean image (all) | 178 | 0 | 5.2 | Figure 10 a |
| Incoherent intensity | 76 | 0 | 7.35 | Figure 10 b, Equation 20 |
| Incoherent intensity PCA | 76 | 0 | 5.35 | Figure 10 c, Section 5.2 |
| Reference PSF subtraction | 38 | 38 | 9.34 | Not shown |
| Reference PCA reduction | 38 | 38 | 7.5 | Not shown |
| Reference PSF subtraction (all) | 178 | 178 | 14.3 | Figure 10 d |
| Reference PCA reduction (all) | 178 | 178 | 19.5 | Not shown |

Table 1:: Summary of reductions. All observations of HD 49197 and reference star HD 48270 were on 22 November 2015.

## 6. Discussion

### 6.1. Potential as a method of speckle suppression

Our results using the central pistoning mirror to measure the phase of the electric field of focal-plane speckles demonstrate that the technique works as theory predicts, and can be used to effectively suppress speckles in regions of interest.

---

[2] intensity matched subtraction was performed at each radial separation from the star, with an annular mask of size 2 $\lambda/D$, using median target and calibrator frames

Compared to methods mentioned in the introduction (speckle nulling, EFC, COFFEE, SCC, SISS), there are some advantages. The technique requires no system model, and has a particularly simple formulation to derive the electric field phase, giving good results with no complicated processing or statistical estimation. One of the reasons for the good performance is the ability to put large phases on the pistoning mirror, unlike in electric field conjugation, where the "probe" steps are typically very small to respect linearity approximations, crosstalk, and stroke limits of the deformable mirror. The phase measurements can be seamlessly integrated into the science measurements, as the reference beam is unobtrusive (at least in this implementation). The extra optical complexity is less than both the self-coherent camera and the synchronous interferometric speckle subtraction techniques, despite its similarity. This approach may also be compatible with a range of coronagraphs, though the design requires optimization around the particular coronagraph design.

However, there drawbacks to this technique. First, the need to precisely determine the focal plane phase gradient caused by the rod decenter and/or optical power requires additional measurements, and the associated uncertainties in these measurements impose a penalty on the phase accuracy. In contrast, techniques using the deformable mirror measure speckles' relative phases quite accurately without requiring a zero-point calibration of any phase gradients in the focal plane. Our measurements of the gradient phase accuracy showed that about 10-15 degrees of accuracy could be reliably obtained; speckle-nulling approaches typically gave better than 5 degrees of accuracy for the same test speckles.

Another disadvantage of this approach is the need for an absolute zeropoint calibration, which means that any drifts in the relative cophasing of the rod and the rest of the optics need to be tracked. In our lab, drifts of $\lambda/4$ occurred over the course of about an hour. At a more unforgiving thermal and mechanical environment, such as a telescope, phase drifts would need to be constantly recalibrated, affecting observing efficiency. For simple linear phase drifts, there are non-invasive ways that could be used, such as generating spots of fixed phase at the extremes of the focal plane using the deformable mirror, and tracking those measured phases actively. More complex changes to the optical path, such as focus and other distortions, would require multiple "probe" spots in the focal plane that would be too obtrusive to science measurements (we used 32 spots over the focal plane for the calibration). EFC, speckle nulling, and COFFEE approaches are immune to these issues as the phase measurement and phase correction device are the same–the deformable mirror. SCC is also probably immune via the static nature of the pinholes. An alternative approach could replace the phase-shifting rod with a full deformable mirror at the second pupil plane; this would remove the error due to co-phasing drift and also allow for full correction of the focal plane, not just phase errors, though the limited stroke of most commercially available deformable mirrors might be an issue. A bimorph deformable mirror could achieve this stroke, but would probably require a custom electrode pattern to achieve a flat surface like the pistoning rod.

## 6.2. Potential as a method of planet-speckle discrimination

The phase-shifting approach shows promise as a method to detect companions. We first point out that none of the issues with the phase stability are relevant for the calculation of the visibilities (Equation 11), as the visibility relies on *differential* phase accuracy between phase steps. The notion of a "phase" for the companion light does not even make sense, as it is incoherent with the starlight. The optical phase thus only needs to be stable on timescales corresponding to a the four readouts of the camera, which is on the order of seconds for many instruments.

Encouragingly, the detection of HD49197b using coherence modulation was straightforward. The observing conditions were average, with a Strehl ratio in the 70-75% range. In normal SDC observing, we perform aggressive point-spread-function sharpening routines using the modified Gerchberg-Saxton algorithm to improve the internal instrumental Strehl ratio to above 90%. With the phase shifter installed, we were unable to do this as the algorithm refused to converge with the unfamiliar pupil geometry. We instead hand-tuned Zernike polynomials to increase the Strehl ratio, but were only able to achieve about 80% in this way. We also had a significant additional component of residual diffraction from the telescope spiders. Normally, we block the spider diffraction using appropriately shaped pupil stops in the infrared camera PHARO, but we were unable to do this as the stops would block light from the phase shifting mirror as well (some excess spider diffraction can be seen in Figure 10). In short, conditions were well below normal quality in this first trial, and the fact that the companion was so easy to detect suggests that the technique is fairly robust.

## 7. Conclusion

We have presented the design, lab and on-sky results from a phase-shifting interferometer installed on the Stellar Double Coronagraph at Palomar observatory. In one application, we demonstrated its use as a focal-plane wavefront sensor with the ability to measure the electric field over the full focal plane, providing a direct measurement of the speckle phases. We used this mode to create a region of high contrast in the focal plane; a "dark hole". Apart from these advantages, we note that measuring and maintaining the absolute phase zeropoint with two separate mirrors caused less robust results in this initial test than conventional speckle nulling, though applications requiring full-frame electric field phase knowledge could benefit from such a device. Moreover, use of a second deformable mirror at the same location would solve this problem.

As a second application, we used the phase-shifter to measure the visibility, rather than the phase, of the speckle field. We described how the visibility can be used to discriminate between coherent and incoherent light in the focal plane, and therefore, between speckles and real companions. We demonstrated this on-sky by detecting the companion HD49197b, both as a region of locally depressed coherence using only data from the target star, and also as a bright point source when combined with coherence data from a reference star. We discussed how this extra information can be used to extract companion photometry using the visibility data, as well as how it can combined with powerful PSF-subtraction techniques like KLIP and LOCI.

We conclude that coherence modulation is a promising technique. The relative ease with which the companion was detected despite poor optical conditions, moderate observing conditions, and minimal data processing or selection strongly hints that extra information from coherence data can be obtained for a small increase in observing complexity. In the current era of extreme-AO systems and upcoming space telescopes such as WFIRST-AFTA, the potential for additional sensitivity gains at high contrasts should not be ignored.

## 8. Acknowledgements

We are very pleased to acknowledge the Palomar Observatory staff for their support. During the course of this work, MB was supported by a NASA Space Technology Research Fellowship, grant NNX13AN42H. Part of this work was carried out at the Jet Propulsion Laboratory, California Institute of Technology, under contract with the National Aeronautics and Space Administration (NASA).

## A. Derivation of Equation 15

Beginning with the error propagation formula, we have

$$\phi_s(x,y) = \tan^{-1}\left[\frac{I_4 - I_2}{I_1 - I_3}\right] \tag{A1}$$

$$\Rightarrow \sigma_\phi^2 = \sum_{i=1}^{4} \sigma_{I_i}^2 \left(\frac{\partial \phi}{\partial I_i}\right)^2 \tag{A2}$$

$$= \sum_{i=1}^{4} I_i \left(\frac{\partial \phi}{\partial I_i}\right)^2 \tag{A3}$$

where the last step follows from the assumption of Poisson statistics; that is, $\sigma_I^2 = I$. Now using the fact that $d/dx \tan^{-1}(x) = 1/(1+x^2)$ and the chain rule, we note that $\left(\frac{\partial \phi}{\partial I_1}\right)^2 = \left(\frac{\partial \phi}{\partial I_3}\right)^2$, and similarly for $I_4$ and $I_2$. We then find

$$\sigma_\phi^2 = (I_1 + I_3)\left(\frac{\partial \phi}{\partial I_1}\right)^2 + (I_2 + I_4)\left(\frac{\partial \phi}{\partial I_2}\right)^2 \tag{A4}$$

$$= (I_s + I_r)\left[\left(\frac{\partial \phi}{\partial I_1}\right)^2 + \left(\frac{\partial \phi}{\partial I_2}\right)^2\right] \tag{A5}$$

The second line follows by direct substitution of the values of $I_1, I_2, I_3, I_4$ from 9. Now we compute the partial derivatives using the chain rule and find

$$\sigma_\phi^2 = (I_s + I_r) \frac{(I_1 - I_3)^2 + (I_4 - I_2)^2}{[(I_1 - I_3)^2 + (I_4 - I_2)^2]^2} \tag{A6}$$

$$= \frac{(I_s + I_r)}{(I_1 - I_3)^2 + (I_4 - I_2)^2} \tag{A7}$$

$$= \frac{(I_s + I_r)}{4 I_s I_r} \tag{A8}$$

## B. Astrometry and photometry of HD 49197b

We performed measurements of the separation, position angle, and contrast ratio of the brown dwarf using the method introduced in Bottom et al. (2015). Briefly, a Markov-chain Monte Carlo algorithm (Foreman-Mackey et al. 2013) tries to create the images of the target frames (with the companion) from the reference frames (without the companion) and an image of the non-coronagraphic point-spread function of the primary, by modifying the PSF position (x and y), brightness, and the reference frame intensity. The four quantities are used to derive the separation, position angle, and magnitude difference. For the absolute position angle, measurements of the astrometric binary HD32022 were used to calibrate the sky rotation angle. For absolute separation, the position of the primary star must be known. This was determined from centroiding the residual diffraction from the telescope spiders, which was not blocked in this observing configuration. This is the dominant source of uncertainty in both separation and position angle, as the algorithm can determine the center of the companion to better than 0.05".

The derived quantities are presented in Table 2, with the posterior distributions given in Figure 11. While the star-to-companion contrast is found to be consistent with previous measurements from Metchev & Hillenbrand (2004) and Serabyn et al. (2009), we detect significant orbital motion over the ∼12 year time baseline. This confirms a statistically insignificant displacement seen by Metchev & Hillenbrand (2004) over a less than two year baseline. The companion is moving inward.

## REFERENCES

Baudoz, P., Boccaletti, A., Baudrand, J., & Rouan, D. 2006, in IAU Colloq. 200: Direct Imaging of Exoplanets: Science & Techniques, ed. C. Aime & F. Vakili, 553–558

Bordé, P. J., & Traub, W. A. 2006, ApJ, 638, 488

Bottom, M., Kuhn, J., Mennesson, B., Mawet, D., Shelton, J. C., Wallace, J. K., & Serabyn, E. 2015, ApJ, 809, 11

| Work | Separation (") | P.A. (deg) | $\Delta K_s$ | Date (JD) |
|---|---|---|---|---|
| Metchev & Hillenbrand (2004) | 0.9499 ± 0.0054 | 78.25 ± 0.40 | 8.22 ± 0.11 | 2452333 |
| Metchev & Hillenbrand (2004) | 0.9475 ± 0.0022 | 77.60 ± 0.25 | 8.22 ± 0.11 | 2452953 |
| Serabyn et al. (2009) | 0.96 ± 0.1 | 77 ±0.2 | 8.2 ± 0.2 | 2453989 |
| This Work | 0.862 ± 0.025 | 76.6 ± 1.8 | 8.18 ± 0.20 | 2457349 |

Table 2::  Astrometric and photometric results from this work and previous work. Uncertainties quoted here for this work are conventional "1 $\sigma$" results, but likely understate the degree of precision of the measurements, as they are dominated by outliers in the sample chain. For comparison, the 25th and 75th quantiles of the data are (0.851, 0.872) arcseconds, (75.95, 77.33) degrees, and (8.160, 8.193) magnitudes.

Bottom, M., et al. 2016, PASP, 128, 075003

Carré, P. 1966, Metrologia, 2, 13

Dekany, R., et al. 2013, ApJ, 776, 130

Foreman-Mackey, D. 2016, The Journal of Open Source Software, 24

Foreman-Mackey, D., Hogg, D. W., Lang, D., & Goodman, J. 2013, PASP, 125, 306

Give'on, A., Kern, B., Shaklan, S., Moody, D. C., & Pueyo, L. 2007, in Proc. SPIE, Vol. 6691, Astronomical Adaptive Optics Systems and Applications III, 66910A

Groff, T. D., Eldorado Riggs, A. J., Kern, B., & Jeremy Kasdin, N. 2015, Journal of Astronomical Telescopes, Instruments, and Systems, 2, 011009

Guyon, O. 2004, ApJ, 615, 562

Hayward, T. L., Brandl, B., Pirger, B., Blacken, C., Gull, G. E., Schoenwald, J., & Houck, J. R. 2001, PASP, 113, 105

Lafrenière, D., Marois, C., Doyon, R., Nadeau, D., & Artigau, É. 2007, ApJ, 660, 770

Malacara, D. 2007, Optical Shop Testing (Wiley Series in Pure and Applied Optics) (Wiley-Interscience)

Marois, C., Lafrenière, D., Doyon, R., Macintosh, B., & Nadeau, D. 2006, ApJ, 641, 556

Martinache, F., et al. 2014, PASP, 126, 565

Mawet, D., Pueyo, L., Carlotti, A., Mennesson, B., Serabyn, E., & Wallace, J. K. 2013, ApJS, 209, 7

Mawet, D., Serabyn, E., Wallace, J. K., & Pueyo, L. 2011, Opt. Lett., 36, 1506

Metchev, S. A., & Hillenbrand, L. A. 2004, ApJ, 617, 1330

Paul, B., Mugnier, L. M., Sauvage, J.-F., Ferrari, M., & Dohlen, K. 2013, Optics Express, 21, 31751

Pueyo, L., Kay, J., Kasdin, N. J., Groff, T., Mc Elwain, M., Give'on, A., & Belikov, R. 2011, ArXiv e-prints

Riggs, A. J. E., Kasdin, N. J., & Groff, T. D. 2016, Journal of Astronomical Telescopes, Instruments, and Systems, 2, 011017

Sauvage, J.-F., Mugnier, L., Paul, B., & Villecroze, R. 2012, Optics Letters, 37, 4808

Serabyn, E., Mawet, D., Bloemhof, E., Haguenauer, P., Mennesson, B., Wallace, K., & Hickey, J. 2009, ApJ, 696, 40

Serabyn, E., Wallace, J. K., & Mawet, D. 2011, Appl. Opt., 50, 5453

Soummer, R., Pueyo, L., & Larkin, J. 2012, ApJ, 755, L28

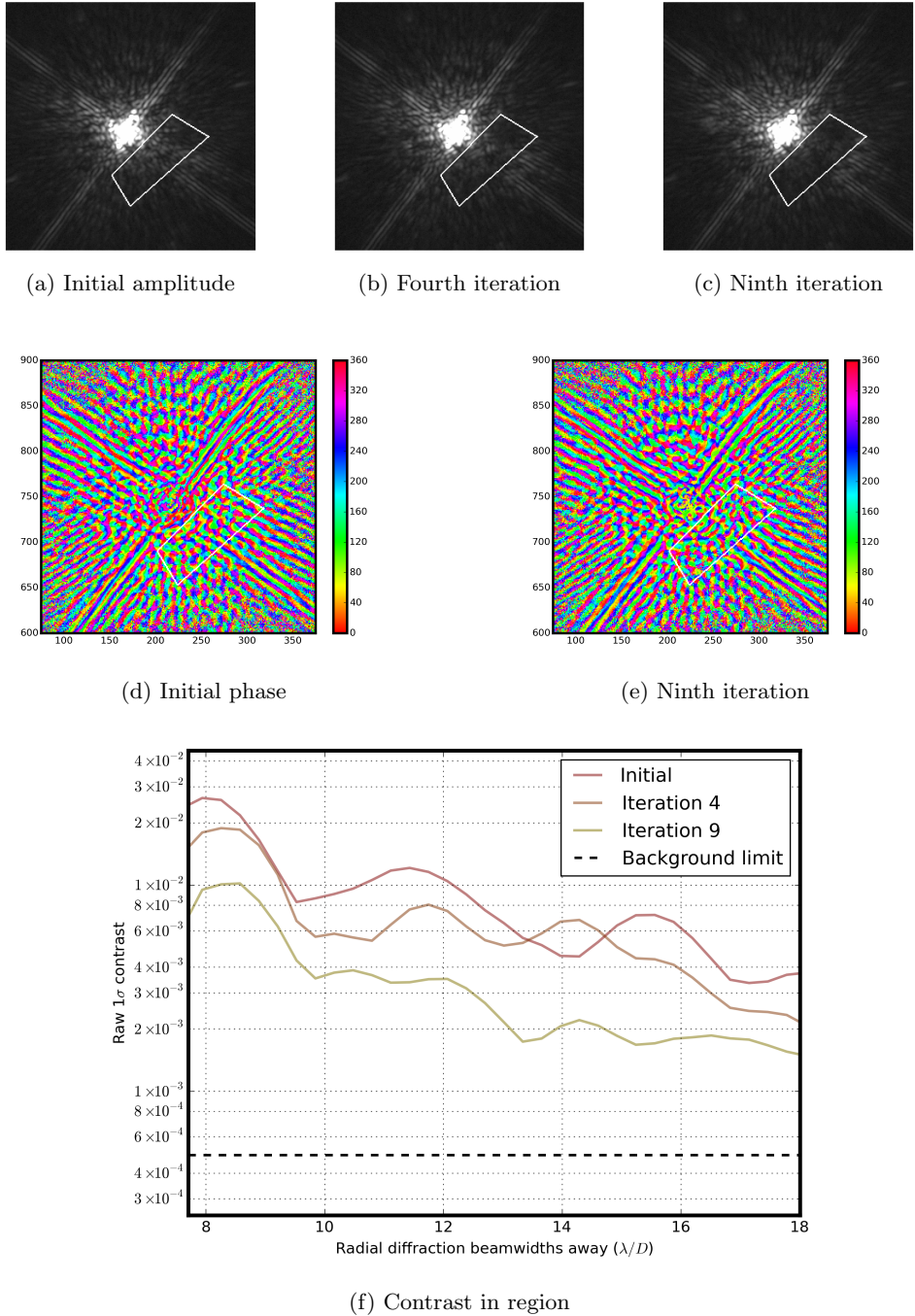

(a) Initial amplitude  (b) Fourth iteration  (c) Ninth iteration

(d) Initial phase  (e) Ninth iteration

(f) Contrast in region

Fig. 9.—: Laboratory demonstration of coronagraphic speckle-suppression using the phase-shifting interferometer. The white outline delineates the region of speckle control. Panels a, b, and c show the intensity after 0, 4, and 9 iterations of speckle removal by the deformable mirror. Panels d and e show the phase measurements of the initial and final light distribution, showing a clear difference in phase. Panel f shows the $1\sigma$ contrast curve measurement of the region after 1, 4, and 9 iterations. The contrast curve is defined in the usual way, with the standard deviation (ie, $1\ \sigma$) of surface brightness at each radial separation being used to generate the curve, and normalized by dividing by the peak flux of the non-coronagraphic PSF (not shown). The preprocessing performed on the data was dark subtraction and flat-fielding.

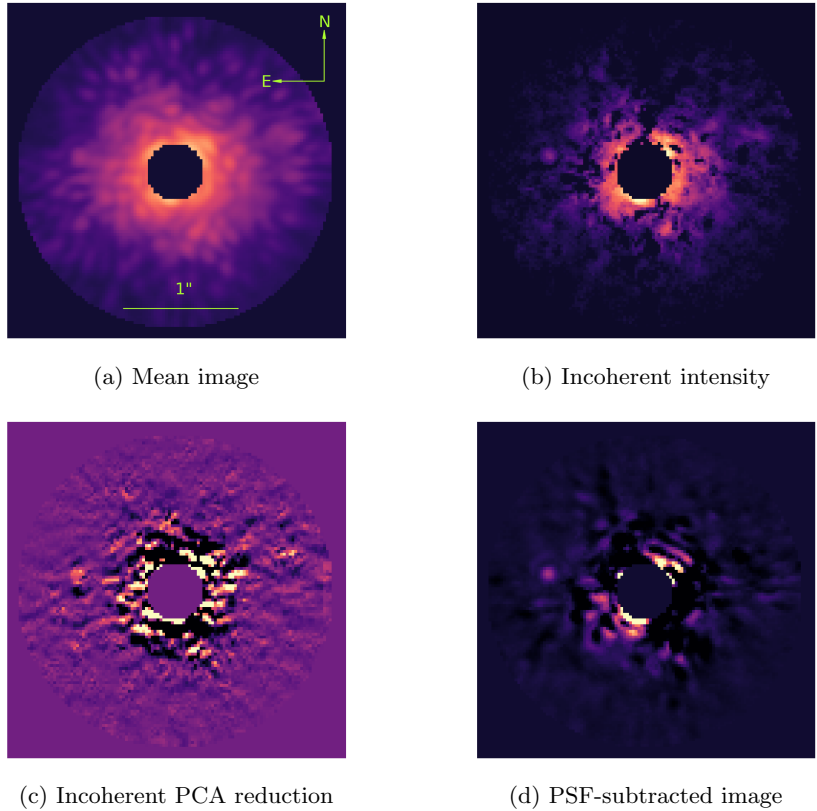

Fig. 10.—: Coherence modulated detection of HD 49197b. (a) The mean of the target frames. The substellar companion is not clearly visible in the image. The companion is present at an SNR of about 5. The stretch is logarithmic. (b) The incoherent intensity map (Equation 20) The companion is easily visible at a position angle of about 275 degrees, and has an SNR of about 7.5. The stretch is also logarithmic. We note that the same data frames were used to compute image (b) as in (a), except combined using the interferometer position to give extra information. No reference star is used (c) Principle components analysis (KLIP) reduction of HD49197b, where the components are generated from only the coherent parts of the image data, as explained in Section 5.2. No reference star is used, and the SNR is 9. The stretch is linear. (d) A conventional PSF-subtracted image of the companion, using a nearby reference star, for comparison. All the data is used, and the SNR is 14. The stretch is linear.

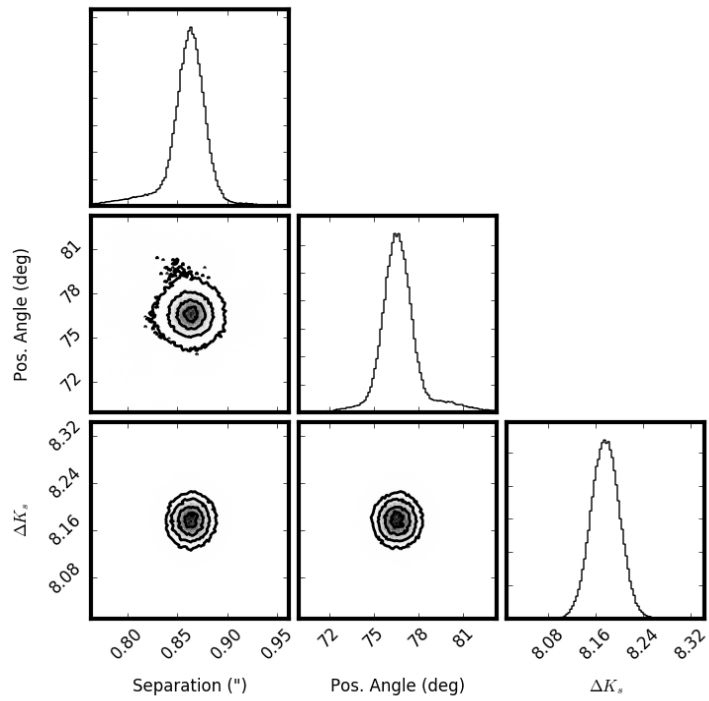

(a)

Fig. 11.—: Posterior distributions of the position angle, separation, and magnitude difference of the companion star. This plot was generated using code presented in Foreman-Mackey (2016).